\documentclass{elsart}
\usepackage{epsfig}
\usepackage{amssymb}

\begin{document}

\begin{frontmatter}

\title{Search for time modulations in the Gallex/GNO solar neutrino data}
\author[a,b,gno]{L. Pandola}
\ead{pandola@lngs.infn.it}
\thanks[gno]{As a member of the GNO Collaboration} 
\address[a]{INFN, Laboratori Nazionali del Gran Sasso, S.S. 17 bis 
km 18+910, I-67010 L'Aquila, Italy}
\address[b]{Universit\`a dell'Aquila, Dipartimento di Fisica, Via 
Vetoio 1, I-67010 L'Aquila, Italy}
\begin{abstract}
The final data of the Gallex and GNO radiochemical 
experiments have been searched for possible time modulations of the 
solar neutrino capture rate using the Lomb-Scargle periodogram and 
the maximum likelihood methods. Both analyses, performed using the  
raw data, do not support the presence of a time variability with 
characteristic periods resembling those of rotation of the solar 
magnetic fields, as previously suggested in the literature by some 
authors.
In this context, the potential sensitivity of the 
Lomb-Scargle method to radiochemical $^{71}$Ga data has also been 
explored with simulated data. This allowed to 
set an exclusion plot in the frequency/amplitude plane from the 
Gallex-GNO dataset. 
\end{abstract}
\end{frontmatter}

% main text
\section{Introduction} \label{one}
\begin{figure}[tb]
\begin{center}
\mbox{\epsfig{file=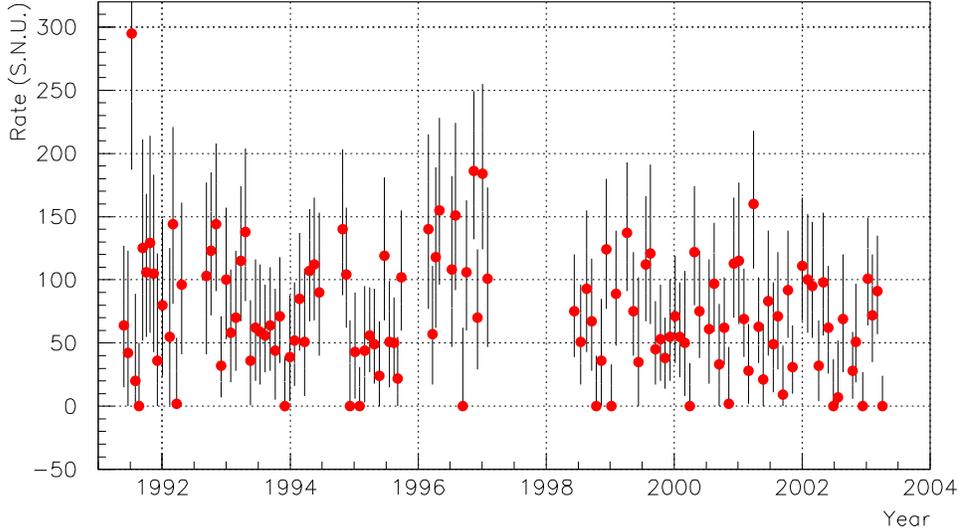,height=9cm}} 
\caption{Measured solar neutrino capture rate (SNU) 
in the 123 solar runs of Gallex/GNO \cite{gno3}.} \label{fig1}
\end{center}
\end{figure}

In recent years it has been suggested \cite{stu1,stu2,stu3,stu4,stu5,mil,stu6} 
that solar neutrino flux measured by Cl and Ga radiochemical experiments and 
by Super-Kamiokande shows a time variability with characteristic periods 
resembling those of rotation of the solar magnetic fields. 
Since this could be a very strong hint for sub-leading 
Resonant-Spin-Flavour-Precession (RSFP) of \emph{pp} neutrinos occurring in the 
solar convection zone \cite{stu7} and for a non-negligible neutrino transition 
magnetic moment, it is of paramount importance to perform an 
extensive search for possible time modulations in the $^{71}$Ga data using 
different methods and exploiting the total information of the available solar runs.\\
The Gallex \cite{gallex} and GNO \cite{gno1,gno2,gno3} radiochemical experiments at 
Laboratori Nazionali del Gran Sasso, 
Italy, monitored the low energy solar neutrino flux using a 30-ton gallium detector 
 via the inverse $\beta$ reaction $^{71}$Ga($\nu_{e}$,e$^{-}$)$^{71}$Ge. 
GNO is the successor project of Gallex, which continuously 
took data between 1992 and 1997, and has been in operation from April 1998 to 
April 2003. 
The gallium detector at Gran Sasso consists 
of 30.3 tons of natural gallium in the form of 103 tons of gallium chloride 
(GaCl$_{3}$) acidic solution. The energy threshold of the 
$^{71}$Ga($\nu_{e}$,e$^{-}$)$^{71}$Ge reaction, 233 keV, is sufficiently low 
to make gallium detectors sensitive to the fundamental \emph{pp} neutrinos.
 Presently 
only radiochemical experiments based on $^{71}$Ga are sensitive to 
these neutrinos. $^{71}$Ge nuclei produced by $\nu_{e}$ interactions 
are unstable and 
transform back to $^{71}$Ga by electron capture, with a life time $\tau$ of 
16.49 days. 
The gallium target is exposed to solar neutrinos for a time 
t$_{exp}$ = t$_{EOE}$-t$_{SOE}$ (SOE = start of exposure, EOE = end of 
exposure) which is typically 3 or 4 weeks. After this, the accumulated 
$^{71}$Ge ($\sim$8 atoms for a 4-week-exposure) is extracted 
from the target solution and converted in germane gas GeH$_{4}$. The germane 
is added with low-activity xenon and is used as counting gas for a 
miniaturized 
proportional counter, which is able to detect the electron capture decays 
of $^{71}$Ge. The interaction rate of solar neutrinos with the target is 
measured, in the hypothesis it is constant in time, by counting the decay 
signals of $^{71}$Ge in the proportional counter. For details about the 
experimental procedure and data analysis see Ref. \cite{gno1}. 
The final result from Gallex (65 solar runs, 1594 days of exposure time) is 
77.5$\pm 6.2 \pm$4.5 SNU.\footnote{1 SNU = 10$^{-36}$ interactions per 
second and per target nucleus.} From April 1998 to April 2003, 58 solar runs, with 
exposure time of 3 or 4 weeks, were successfully performed within the GNO 
project, for a overall time of 1713 days \cite{gno1,gno2,gno3}: the resulting solar 
neutrino interaction rate is 62.9$^{+5.5}_{-5.3} \pm$2.5 SNU. The combined 
Gallex/GNO rate (123 solar runs, 3307 days of exposure time) is 
69.3$\pm 4.1 \pm$3.6 SNU. Results from the single solar runs are displayed 
in Fig. \ref{fig1}.\\
In this paper, possible significant time periodicities in the 
solar neutrino data are looked for using the Lomb-Scargle periodogram method 
(which has already been used in the literature for this kind of analysis); 
the potential sensitivity is also tested with simulated data sets. 
The search for time periodicities is then repeated by applying the maximum 
likelihood 
method to the candidate $^{71}$Ge decays in the single solar runs;  
the constant solar neutrino rate hypothesis is tested and 
a likelihood spectrum is derived. \\ 
\nopagebreak
\section{Lomb-Scargle periodogram method} \label{two}
\subsection{Analysis of the solar runs}
\begin{figure}[tb]
\begin{center}
\mbox{\epsfig{file=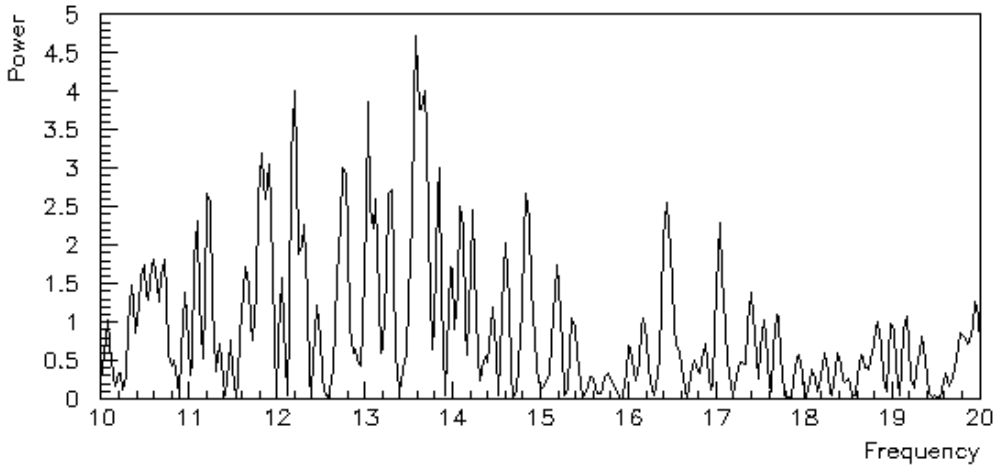,width=14cm}} 
\caption{Lomb-Scargle periodogram for the 84 solar runs of Gallex and GNO I 
\cite{gallex,gno1} in the frequency range $10-20$y$^{-1}$. 
This plot corresponds to Fig. 2 of Ref. \cite{stu2}}  \label{sturrock}
\end{center}
\end{figure}

\begin{figure}[tb]
\begin{center}
\mbox{\epsfig{file=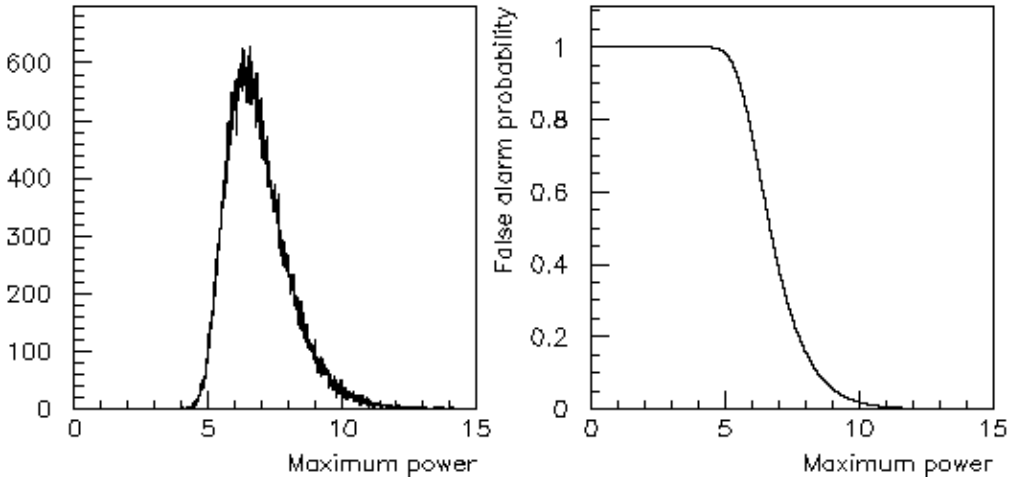,width=14cm}} 
\caption{Distribution of maximum power $z_{max}$ (left) and false-alarm 
probability (right) for the null hypothesis (i.e. constant signal with 
random noise), assuming $N=123$ data with times $\{ t_{i} \}$ as 
in the real solar runs and frequency in the range 
[$\nu_{min}$,$\nu_{max}$].}  \label{MCplot}
\end{center}
\end{figure}

The search for periodicities in the Gallex/GNO solar neutrino data can be performed 
using the Lomb-Scargle periodogram method (see  
Ref. \cite{stu2,mil,sk-web}). Given a set of $N$ data points 
$ h_{i}  \equiv h(t_{i})$ at the times $\{ t_{i} \}$, $i = 1 \dots N$, 
with mean and variance 
$\bar{h}$ and $\sigma^{2}$ respectively, the Lomb-Scargle periodogram 
is defined  
as \cite{lomb,scargle}
\begin{equation}
z(\nu)=\frac{1}{2 \sigma^{2}} \Big{\{} 
\frac{[\sum_{j}(h_{j}-\bar{h}) \cos \omega (t_{j}-\theta)]^{2}}
{\sum_{j}\cos^{2} \omega(t_{j}-\theta)} + 
\frac{[\sum_{j}(h_{j}-\bar{h}) \sin \omega (t_{j}-\theta)]^{2}}
{\sum_{j}\sin^{2} \omega(t_{j}-\theta)} \Big{\}}, \label{periodogram}  
\end{equation}
where $\omega  \ = \ 2 \pi \nu$ is the angular speed and $\nu$ is the 
frequency; the offset $\theta$ is defined by the relation 
\begin{equation}
\tan(2 \omega \theta) \ = \ \frac{\sum_{j} \sin (2 \omega t_{j})}
{\sum_{j} \cos (2 \omega t_{j})}
\end{equation} 
and makes the periodogram $z(\nu)$, which is also called power, independent 
of shifting all the times 
$\{ t_{i} \}$ by 
any constant \cite{scargle}. \\
Following the procedure used in Ref. \cite{sk-web}, the 
periodogram for the Gallex/GNO data presented in Ref. \cite{stu2} is first 
reproduced. 
Fig. \ref{sturrock}
shows the Lomb-Scargle periodogram obtained from the analysis of the first 84 
Gallex/GNO solar runs \cite{gallex,gno1}, that can be directly compared with Fig. 
2 at the page 1369 of Ref. \cite{stu2}.    
\nopagebreak
The most relevant features of the original periodogram are 
correctly reproduced: the peaks (and in particular the main one, at frequency 
$\sim$ 13.7 y$^{-1}$) are approximately in the same position and the vertical 
scale of power, which fixes the statistical significance of the peaks themselves, 
is exactly equal. This level of agreement is completely satisfactory for 
the present testing purposes\footnote{Minor differences in the positions or in the powers of 
the single peaks, that are probably related to different used 
approximations, are not significant in this context.}. \\
After this test of the Lomb-Scargle code, the same analysis has been 
extended to the 
whole set of Gallex and GNO data \cite{gallex,gno1,gno2,gno3}. 
As in the preliminary test of Fig. \ref{sturrock}, the results 
$\{ h_{i} \}$ of the $N=123$ solar runs are conventionally 
referred to the end-of-exposure  $\{ t^{i}_{EOE} \}$ 
times%\footnote{Since the Lomb-Scargle periodogram 
%$z(\nu)$ is invariant for time shifts and the 
%exposure times for the solar runs are not very different, this 
%is almost equivalent to refer the data to the the mean 
%$^{71}$Ge production time, which can be calculated in the 
%hypothesis of constant neutrino rate.} 
 \cite{stu2}. 
In order to test the region of major astrophysical interest, 
the frequency range taken into account spans from
\begin{equation}  
\nu_{min} = \frac{1}{2(t_{N}-t_{1})} = 0.04 \ \textrm{y}^{-1} \quad 
\textrm{to} \quad \nu_{max} = 5 \cdot \nu_{cr} = 26 \ \textrm{y}^{-1}, 
\end{equation}
where
$\nu_{cr}=\frac{N}{2(t_{N}-t_{1})} = $ 5.19 y$^{-1}$ is the average 
Nyquist frequency.
In the periodogram analysis all the data points $\{ h_{j} \}$ 
are used with equal statistical weights and 
are corrected to take into account the 
expected $1/d^{2}$ geometrical modulation of the neutrino flux, which is 
due to the variation of the Sun-Earth distance during the year. 
If $z_{max}$ is the highest peak found in the periodogram, one has to 
extablish if this is compatible with statistical fluctuations, i.e. one 
has to test the \emph{null hypothesis} $H_{0}$ of a 
constant neutrino signal with random noise.
The probability 
$P(>z_{max} | H_{0})$
that statistical fluctuations give a peak in the periodogram 
larger than $z_{max}$ 
%, if some $M$ independent frequencies are scanned, is 
%\cite{scargle,nr}
%\begin{equation}
%P(>z_{max} | H_{0}) \ = \ 1-(1-e^{-z_{max}})^{M}; \label{null}
%\end{equation}
%this 
is the false-alarm probability of the null hypothesis and a small value 
of $P(>z_{max} | H_{0})$ indicates an highly significant periodic signal. 
Following the procedure of Ref. \cite{nr}, the false-alarm probability has been 
estimated by generating 100,000 Monte Carlo data 
samples $\{ h_{i} \}$, $i=1 \dots 
123$, in the assumption of the hypothesis 
$H_{0}$. As the probability depends also 
on the detailed time spacing of the data, the $\{ h_{i} \}$ have been  
referred to the same times $\{ t_{i} \}$ as the real runs. The 
Lomb-Scargle method is then applied to each data sample, to get the 
spectrum $z(\nu)$ and the maximum power $z_{max}$. The resulting maximum 
power distribution and the false-alarm probability are displayed in  
Fig. \ref{MCplot}; the $z_{max}$ power values 
corresponding to 50$\%$, 5$\%$, 1$\%$ and 0.1$\%$ false-alarm probabilities 
are respectively 6.69, 9.12, 10.57 and 12.55.\footnote{The 50$\%$ 
false-alarm level, $z_{max}$=6.69, corresponds to the average 
value of $z_{max}$ expected from a null hypothesis signal.} \\
\nopagebreak
\nopagebreak
\begin{figure}[tb]
\begin{center}
\mbox{\epsfig{file=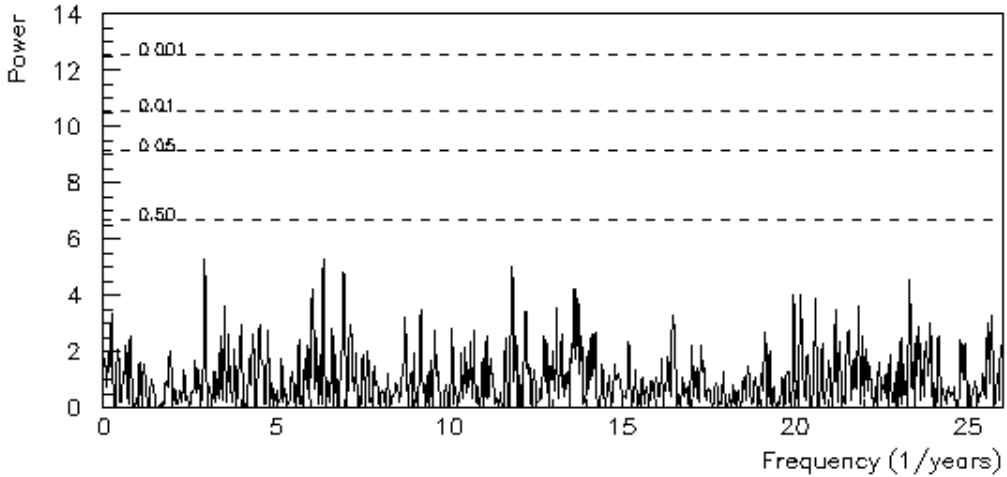,width=14cm}} 
\caption{Lomb-Scargle periodogram for the 123 solar runs of Gallex/GNO in 
the frequency range [$\nu_{min}$,$\nu_{max}$]; dashed lines are 
the 50$\%$, 5$\%$, 1$\%$ and 0.1$\%$ false-alarm levels calculated from Fig. \ref{MCplot}. } 
 \label{dataperiodogram}
\end{center}
\end{figure}
The Lomb-Scargle periodogram obtained from the analysis of the 123 solar runs 
of Gallex and GNO is shown in Fig. \ref{dataperiodogram}. The highest peak in 
the interval [$\nu_{min}$,$\nu_{max}$] is found at $\nu_{max} = 6.37 \pm 0.04$ y$^{-1}$; 
the corresponding power is $z_{max} = 5.26$ and the false-alarm probability for the 
null hypothesis is $\sim$95$\%$. The Lomb-Scargle periodogram for Gallex/GNO solar 
neutrino data is hence statistically consistent with the expectation of a constant 
interaction rate and it does not contain any hint of time modulation. If this same 
analysis is repeated for the 58 solar runs of GNO only, the highest peak is found at
the frequency $\nu_{max} = 2.9 \pm 0.1$ y$^{-1}$, with false-alarm probability 
$\sim$55$\%$.
\subsection{Test of the sensitivity with simulated data} \label{mctest}
\begin{figure}[tp]
\begin{center}
\mbox{\epsfig{file=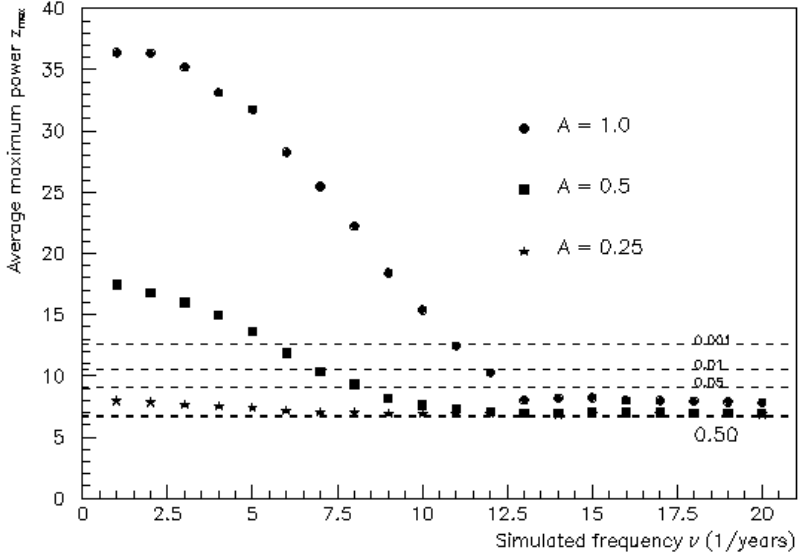,width=11cm}} 
\caption{Maximum power $z_{max}(A,\nu)$ vs. frequency 
obtained from the Lomb-Scargle 
analysis of simulated data, in the hypothesis time modulation with 
frequency $\nu$ and amplitude $A=1, \frac{1}{2}$ and  $\frac{1}{4}$.  
The statistical 
error on the single run is assumed to be 40 SNU.
Dashed lines are 
the 50$\%$, 5$\%$, 1$\%$ and 0.1$\%$ false-alarm levels.}. \label{finalplot} 
\end{center}
\end{figure}

In order to test the actual sensitivity of this kind of data to possible time 
modulations, Monte Carlo repetitions $\{ h_{i} \}$ of the Gallex/GNO 
experiments have been generated, in 
the hypothesis that the $^{71}$Ge neutrino-induced production rate is modulated 
in time as
\begin{equation}
n (t) \ = \ n_{0} [1+ A \sin (2 \pi \nu t + \varphi)], \label{modul} 
\end{equation}
with 0 $\le A \le$ 1.
The mean production rate $n_{0}$ is assumed to be 
$n_{0}$ = 0.63 atoms/day = 70 SNU, close to the experimental value. 
The statistical fluctuations on the single runs can be approximated with 
a Gaussian having rms $\sigma = \sim$40 SNU.\footnote{The actual distribution 
of the results of single runs is not exactly Gaussian, because of the asymmetry 
introduced by Poissonian  effects. However, it has been verified that the 
assumption of Gaussian fluctuations does not affect the final result in this 
particular application.}
The search for periodicities in the simulated data has been then performed 
using the Lomb-Scargle periodogram method in the frequency range 
$[\nu_{min},\nu_{max}]$ defined above. Being the Lomb-Scargle periodogram 
invariant for time translations \cite{scargle}, it can be assumed without 
loss of generality that $\varphi=0$, since this is simply equivalent to shift the 
time origin. For each data set, the maximum power $z_{max}(A,\nu)$ 
was evaluated.
% the procedure was repeated 1,000 times, with fixed $\nu$ and $A$, 
%in order to get rid of statical fluctuations on $z_{max}$. 
Fig. \ref{finalplot}  
shows the $z_{max}(A,\nu)$ for frequencies ranging in 1-20 y$^{-1}$ and for 
amplitudes $A=1, \frac{1}{2}$ and  $\frac{1}{4}$. 
\nopagebreak
It can be seen from Fig. \ref{finalplot} that the 
method clearly indentifies the time modulation if the frequency is low and 
the amplitude is large enough. However, even for $A$=100\%, 
the sensitivity rapidly decreases for frequencies $\nu$ higher than 
$\sim 12$ y$^{-1}$: in this case, the modulation signal cannot be 
statistically 
distinguished from a constant signal 
with Gaussian fluctuation, i.e. it is impossible to reject the null 
hypothesis using this kind of data and analysis. \\
\subsection{Exclusion plot}
\begin{figure}[tp]
\begin{center}
\mbox{\epsfig{file=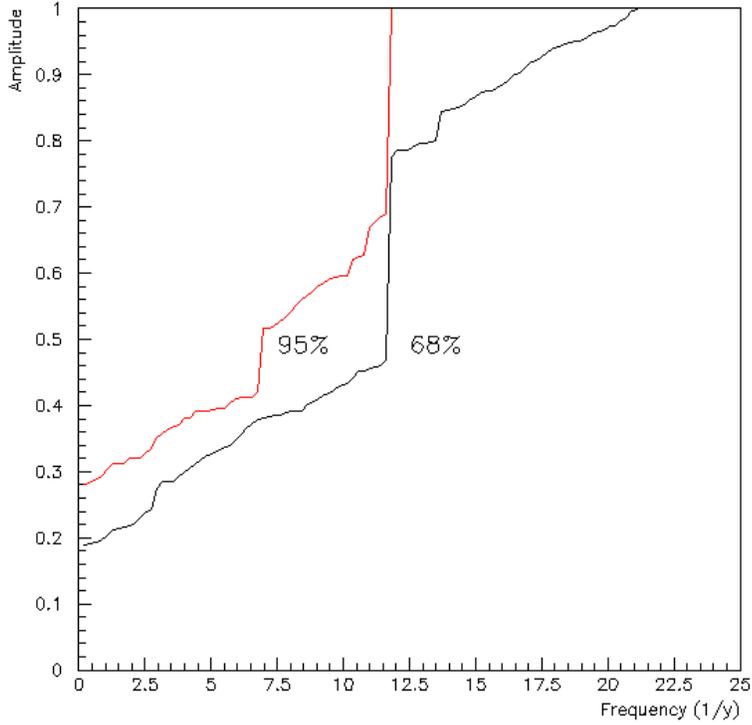,width=11cm}} 
% exclusionplot.gif contiene il 90%
\caption{68\% and 95\% CL exclusion plots in the frequency/amplitude plane 
derived from 
the Lomb-Scargle analysis of GALLEX/GNO data}. \label{exclusionplot} 
\end{center}
\end{figure}

Since no positive modulation signal has been found with the periodogram 
analysis of the Gallex/GNO dataset in the region of sensitivity derived in 
Sect. \ref{mctest}, it is possible to set an exclusion plot. In the assumption of 
time modulation of Eq. (\ref{modul}), the experimental data allow to rule out (see 
Fig. \ref{exclusionplot}) variations of the solar neutrino flux with low 
frequency and large amplitude. As expected, there is a rapid loss of sensitivity 
for $\nu \gtrsim \frac{1}{t_{exp}}$.
\nopagebreak
\section{Maximum likelihood analysis} \label{three}
\begin{figure}[tp]
\begin{center}
\mbox{\epsfig{file=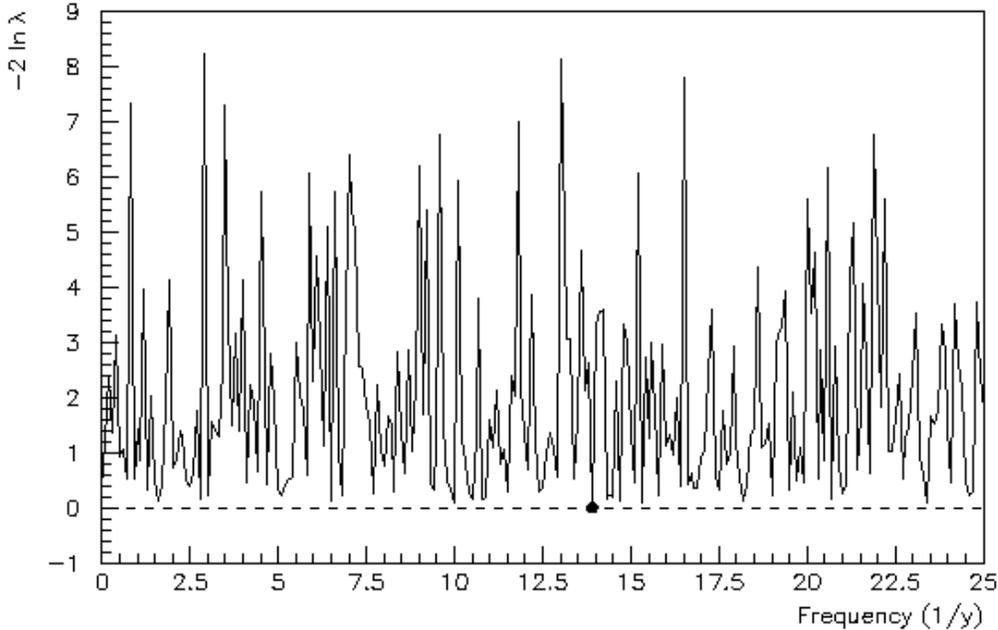,width=14cm}} 
\caption{Likelihood spectrum $-2 \ln \lambda (\nu)$ from the Gallex/GNO dataset vs. 
frequency. All the other parameters are left free in the fit. The black dot marks the 
best-fit point.} \label{mlscan} 
\end{center}
\end{figure}

An other option is to apply the maximum likelihood method to the time 
list of the $^{71}$Ge candidate events in the solar runs, including (on 
the contrary of what happens in the standard Gallex/GNO analysis) a 
time-modulated term in the likelihood function.
The general maximum 
likelihood method described in Ref. \cite{cleveland} can then be applied  
to the case of a production rate varying as in 
Eq. (\ref{modul}). 
In this condition the mean number of $^{71}$Ge atoms produced in 
the target 
depends on the times of start and end of exposure (and not only from the total exposure 
time) and can be calculated as 
\begin{equation}
N_{71} \ = \ \int_{t_{SOE}}^{t_{EOE}} \  
n(t;n_{0},A,\nu,\varphi) \ e^{- \lambda_{71} (t_{exp}-t)} \ dt
\end{equation} 
(see Eq. 12 of Ref. \cite{cleveland}), 
where $\lambda_{71}$ is the decay rate of $^{71}$Ge.
It is hence possible to estimate from the solar run data the best-fit values 
for the four free parameters $n_{0}$, $A$, $\nu$ and $\varphi$ of 
Eq. (\ref{modul}). \\
A known time modulation is actually expected in the solar neutrino data because of 
the seasonal variation of the Sun-Earth distance. Since the excentricity of the 
Earth's orbit is $\epsilon$=0.0167 and the perihelion is on January 4$^{th}$,  
the expected parameters for the 1/$d^{2}$ geometrical modulation in the solar 
neutrino flux are: $A \sim 2 \varepsilon = 0.033$, $\nu$ = 1 y$^{-1}$ and 
$\varphi$= 3.81 rad\footnote{The time origin is conventionally 
set to May 14$^{th}$, 1991, so that $t_{SOE}^{(1)}=0$.}. Although, at least at 
the present stage, the $^{71}$Ga data have not enough sensitivity to 
reveal such a low-amplitude modulation 
($A \cdot n_{0} \sim$ 2.3 SNU, to be compared with the overall Gallex/GNO error, 
$\sim$ 5.5 SNU),  
this is anyway explored using the maximum likelihood method: the frequency and 
phase are fixed to the known values reported above while 
the amplitude is left as a 
free parameter. 
The best-fit coming from the combined analysis of the 123 Gallex/GNO 
solar runs is $n_{0}$=69.3$\pm$5.9 SNU and $A$=0.01$\pm$0.11. The 
amplitude is well consistent with what expected, but it is 
nevertheless consistent with zero, confirming that the presently reached 
sensitivity is not good enough to test this modulation\footnote{The exposure 
needed to reduce the statistical error to $A \cdot n_{0} \sim$ 2.3 SNU would 
be of $\sim$ 700 ton$\cdot$y, to be compared with 270 ton$\cdot$y of Gallex and 
GNO. Moreover the present systematic error, which is 2.5 SNU for GNO, should 
be further reduced in order to reach the required level of sensitivity.}. \\ 
The data analysis was then repeated including two modulation terms in the 
likelihood: one, 
whose parameters are fixed, to account for the annual modulation 
and the other, whose parameters are left free, for other possible unknown periodicities. 
In this case the best-fit point for the 123 solar runs is (statistical errors only)
$n_{0}$=69.1$\pm$6.1 SNU, $A$=0.57$\pm$0.38, $\nu$=13.9$\pm$0.4 y$^{-1}$ (in 
good agreement with the frequency 13.59 y$^{-1}$ reported in Ref. \cite{stu5}) 
and $\varphi$= -0.3$\pm$1.0 rad. 
%However, since the likelihood varies very rapidly 
%with the frequency, the best-fit is not really significant from the point of view of 
%statistics.\\   
%
The hypothesis of modulated solar neutrino flux can be compared with the usual 
null hypothesis of stationary flux (with the geometrical correction) using the 
likelihood ratio test \cite{eadie}. If the constraint $A=0$ is applied (i.e. one 
makes the hypothesis of a constant neutrino flux) the best-fit value is of course 
$n_{0} \ = \ 69.3 \pm 4.1$ SNU (statistical error only).
If 
\begin{equation}
\lambda \ = \ \frac{L_{max}(A=0)}{L_{max}}
\end{equation}
is the likelihood ratio, the statistics $-2 \ln \lambda$ has the same distribution 
of $\chi^{2}(3)$ \cite{eadie}; in the present case, $-2 \ln \lambda = 3.40$ and the 
corresponding $p$-value is 33.4$\%$. The experimental data are hence statistically 
consistent with the hypothesis of a time-constant neutrino interaction rate, 
in agreement with the results of the Lomb-Scargle analysis. \\
In Fig. \ref{mlscan} it is shown the likelihood spectrum $-2 \ln \lambda$ 
obtained fixing 
the frequency and fitting all the other parameters from the data. It can be seen 
from the plot that the absolute maximum of the likelihood is not much 
larger than the other local 
maxima: therefore, no clear modulation frequencies can be derived from the data. 
\nopagebreak
Using the same likelihood ratio test, it is possible to compare the hypothesis of 
stationary rate with the hypothesis of a time-modulated rate with frequency 
13.59 y$^{-1}$ \cite{stu5}:  in this case $-2 \ln \lambda$=3.28 (2 d.o.f.) and 
the corresponding $p$-value is 19.4$\%$.  
\section{Conclusions} \label{four}
The experimental data of the Gallex and GNO solar neutrino experiments (123 solar runs, 
for a total exposure time of 3307 days) have been searched for possible time modulations 
in the capture rate using the Lomb-Scargle periodogram and the maximum likelihood 
methods. For the latter analysis, the time 
list of candidate $^{71}$Ge events in the single runs has been taken into account. 
In both cases no statistical evidence of time variations has been found: given the 
present sensitivity, the results are fully consistent with the expectation of a 
constant solar neutrino capture rate 
over the data taking period ($\gtrsim$ 10 y). 
Though this fact does not automatically exclude other 
hypotheses that predict a time-dependent rate, modulations with low frequency and large 
amplitude can be ruled out.  
\section{Acknowledgements} 
The author acknowledges the permission of the
Gallex Collaboration to access the raw Gallex data for the maximum
likelihood analysis of Sect. \ref{three}. 
The author thanks in particular E.~Bellotti, R.~Bernabei, B.~Cleveland and 
F.~Vissani for very 
useful suggestions and discussions. P.~Belli, C.~Cattadori, T.~Kirsten   
and N.~Ferrari are also kindly acknowledged for their useful feedback.

\end{document}